\pdfoutput=1
\documentclass[manuscript,screen,nonacm,10pt]{acmart}

\acmYear{2022}
\begin{document}

\title{A Hazard Analysis Framework for Code Synthesis Large Language Models}
\author{Heidy Khlaaf}
\authornote{Primary Authors.}
\authornote{Work done while at OpenAI.}
\email{hello@heidyk.com}
\affiliation{UK}

\author{Pamela Mishkin}
\authornotemark[1]
\email{pamela@openai.com}
\affiliation{%
  \institution{OpenAI}
  \country{USA}
}
\email{pamela@openai.com}

\author{Joshua Achiam}
\affiliation{%
  \institution{OpenAI}
  \country{USA}}
\email{jachiam@openai.com}

\author{Gretchen Krueger}
\affiliation{%
  \institution{OpenAI}
  \country{USA}
}
\email{gretchen@openai.com}

\author{Miles Brundage}
\affiliation{%
  \institution{OpenAI}
  \country{USA}
}
\email{miles@openai.com}
\authorsaddresses{}
\renewcommand{\shortauthors}{Khlaaf and Mishkin, et al.}

\begin{abstract}
  Codex, a large language model (LLM) trained on a variety of codebases, exceeds the previous state of the art in its capacity to synthesize and generate code. Although Codex provides a plethora of benefits, models that may generate code on such scale have significant limitations, alignment problems, the potential to be misused, and the possibility to increase the rate of progress in technical fields that may themselves have destabilizing impacts or have misuse potential. Yet such safety impacts are not yet known or remain to be explored. In this paper, we outline a hazard analysis framework constructed at OpenAI to uncover hazards or safety risks that the deployment of models like Codex may impose technically, socially, politically, and economically. The analysis is informed by a novel evaluation framework that determines the capacity of advanced code generation techniques against the complexity and expressivity of specification prompts, and their capability to understand and execute them relative to human ability.
\end{abstract}

\maketitle

\section{Introduction}
Neural network models that generate code have the potential to be useful in a range of ways, from onboarding users to new codebases, to reducing context switching for experienced coders, to education and exploration. However, such models have significant limitations, alignment problems, the potential to be misused, and the potential to increase the rate of progress in technical fields that may themselves have destabilizing impacts or have misuse potential. As discussed in \cite{chen2021evaluating}, Codex, a GPT language model finetuned on publicly available code from GitHub, poses significant safety challenges along these lines. 
%
% This paper addresses the safety framework undertaken to ensure that Codex has been deemed safe so far as is reasonably practicable, and the lessons learned from our analyses. 
%
This paper describes the safety framework undertaken at OpenAI to assess risks related to the deployment of code synthesis large language models (LLMs)\footnote{Note that our analysis targets (and our term ``code synthesis LLM" refers to) language models that have specifically been trained to generate code (e.g. by fine-tuning a base LLM on pure code), rather than language models that only incidentally generate code due to being trained on a small amount of code as part of a larger, diverse dataset, though there is not a hard and fast distinction between these categories.} like Codex, assuming that they are made available to end users through systems like an API or the Github Copilot assistant. We focus primarily on assessing the generative capabilities of these models and risks attached to generative uses, though these models can be used for a variety other tasks such as classification.
Although we initially developed this framework to study Codex specifically, the increasing prevalence of LLMs and their applications to code synthesis makes our approach of general interest in the safe development and deployment of code synthesis LLMs.

%In order to better understand Codex’s limitations and safety implications, first we developed an evaluation framework that code synthesis models’ interpretations and implementations can be measured against. 
In order to better understand Codex’s limitations and safety implications, we first developed an evaluation framework for assessing model capabilities. 
Our capabilities evaluation framework includes a set of qualitative attributes and test problems aiming to measure the extent to which models can generate code meeting increasingly complex and higher-level specifications. Evaluating the capabilities of code synthesis and generation is not a novel problem and has been explored in both the Machine Learning (ML) \cite{xu2021ide} and Synthesis \cite{helmuth2015general, pantridge2017difficulty,gecco15}  communities. However, given the limited capabilities of code generation thus far, evaluation metrics have assumed relatively “simple” function or module-level problems requiring only a range of data types, outputs, and control structures to be demonstrated. Furthermore, these evaluations have not considered safety implications (e.g., fairness, bias, discrimination, etc.) of these technologies’ misuse. Our evaluation framework is appropriate to use for large language models generating code, though a present limitation is that it requires significant effort by a human expert to interpret and classify model outputs.

%An evaluation of the aforementioned metrics are then used to inform a hazard analysis specifically tailored for the evaluation of large language models in order to evaluate the safety concerns of using Codex in a generative capacity. The analysis focuses on identifying risk factors \cite{leveson2019, dod2012mil} with the potential to cause harm against a set of novel losses that form the foundations of addressing safety efforts of large language models. 
The capabilities evaluation informs a hazard analysis specifically tailored for large language models that generate code, like Codex. We describe how to perform our hazard analysis in general, and demonstrate with the hazard analysis for an API system permitting end users to make generative calls to Codex. The analysis focuses on identifying risk factors \cite{leveson2019, dod2012mil} with the potential to cause harm against a set of novel harms intended to form the foundations of safety efforts for general-purpose large language models.

Hazard analysis is a technique typically used in safety-critical systems that serves to collect and interpret information on hazards and conditions that lead to their presence, to determine significant risks that lead to unsafe behavior. A hazard analysis thus informs our risk assessment, in which risks are assessed within the context of the probability and severity\footnote{In addition to probability and severity, distribution was also considered in scenarios in which the harms resulting from a given hazard could be concentrated, e.g., on a specific demographic group.} of the hazard becoming reality. However, unlike traditional safety-critical systems, the potential safety hazards, failure modes, and risks of ML models and their applications are often poorly understood, making a hazard analysis challenging. Hence we emphasize, as a starting point for our hazard analysis, a novel methodical evaluation of the system's capabilities.

In Section \ref{Evaluation}, we define the set of qualitative metrics that aim to benchmark increasingly complex or higher-level specifications to measure the capabilities of advancing code synthesis and generation methodologies. We propose adapting attributes or metrics traditionally utilized to measure the expressivity and complexity of formal specifications to natural language prompts. We then construct a set of preliminary benchmarks given the defined attributes, and evaluate the Codex model against them. We cover the details of our hazard analysis and risk assessment process tailored towards language models in Section \ref{HazardAnalysis}, followed by the highest priority risks identified in Section \ref{Risk}. In Section \ref{HazardControls} we propose a set of mitigation strategies that would alleviate the risks for Codex, followed by next steps and conclusive remarks in Section \ref{Conclusion}.

\section{Evaluation of Capabilities of Language Model-based Code Generation}\label{Evaluation}
Evaluating the capabilities of code synthesis and generation is not a novel problem and has been explored in both the ML \cite{xu2021ide} and Synthesis \cite{helmuth2015general, pantridge2017difficulty} communities. Previously, researchers have recommended the use of existing metrics such as McCabe Cyclomatic Complexity (CC)\cite{xu2021ide}, which provides a quantitative measure of the number of linearly independent paths in a program. However, CC only aims to provide a correlation with the number of defects or bugs that may be within a program. That is, the more branching and execution paths possible, the more likely that a developer may have had a lapse in judgment and thus introduced program defects. This is not a metric for assessing human-level capabilities, as depending on the complexity of the task at hand and the experience of a developer, the CC may be higher or lower.

Another existing metric such as algorithmic complexity is a measure of how long the produced algorithm would take to complete given an input of size n. A scalable algorithm would ideally compute the result within a finite and practical time bound even for large values of n. However, there is no direct correlation between algorithmic complexity and human capabilities, and it is difficult to assess an algorithm without considering the problem at hand. That is, synthesis and generation metrics have largely concentrated on analyzing the correctness and complexity of the code output rather than the expressivity and complexity of the specification itself. Yet, evaluating the output of synthesized code is moot if there is no specification that it can be measured against. Indeed, the Synthesis and automatic programming community \cite{o2020automatic} have recently called for principled benchmarks and grand challenge problems to be made in order to adopt a scientifically rigorous approach against which to compare synthesis methodologies.

\textbf{We should be evaluating generation and synthesis models against the complexity and expressivity of specification prompts and their capability to understand and execute them if we wish to understand their performance relative to human ability.} The remainder of this section thus describes challenges with specification metrics, and recommends a set of qualitative metrics or attributes against which specification prompts can be measured.

\subsection{Specification Complexity and Expressivity}
One of the challenges of traditional code generation and synthesis is that it relies on the assumption that user intent is captured sufficiently enough such that the accuracy and synthesis of a methodology are not compromised. However, from a developer’s standpoint, natural languages are very expressive yet very imprecise and ambiguities are likely to occur, especially among those not versed in defining system requirements. A significant barrier to synthesis is the degree of ambiguity for increasingly higher-level specifications regarding the intent of the system. This has led the majority of synthesis methodologies to tackle only tightly specified, constrained problem instances or narrow tasks requiring much smaller datasets (e.g., string manipulation by FlashFill \cite{flashfill}).

Contrarily, many formal specification languages are both expressive and precise. For example, temporal logic bridges the expression and precision gap by providing a single logical system for describing the program at any level of abstraction, from the highest-level specification to the programming-language implementation. A statement about the program at one level is a meaningful statement about any lower level. However, using formal specifications as basis for synthesis methodologies is impractical, as is done in \cite{blkadek2018counterexample}, if we wish to bring the power of synthesis and code generation to everyday development and productivity. Additionally, Codex synthesizes Python, Javascript, Typescript, and Ruby code, all of which are not amenable to verification \cite{brundage2020toward}; thus it would be difficult to leverage formal specification and verification techniques to evaluate the generated output. Indeed, formal specifications are typically only defined as in scope for safety-critical systems and the barrier of entry is high for everyday developers.

Given the ambiguity of natural language specifications, the challenge arises in how to define an appropriate set of benchmarks with increasingly complex and higher-level specifications to measure the capabilities of advancing code synthesis and generation methodologies. We propose adapting attributes utilized to measure the expressivity and complexity of formal specifications to natural language prompts. This entails evaluating the ability to reason over computations and states at different levels of abstractions as a base metric for complexity and expressivity (e.g., variable dependencies, inter-procedural reasoning, computational interleavings, etc.). Given that this is a complex issue with many layers, we assume that a user is versed and familiar with defining system requirements as suggested by the requirements engineering community \cite{mavin2010big}. Below, we define what we mean by “high-level” specifications and “complex” computational and state reasoning, and define corresponding attributes for each.\footnote{We make this assumption for the purpose of understanding the full extent of Codex's problem-solving capabilities, though in the hazard analysis we also consider potential risks related to system use by inexperienced users.}

\subsection{Specification Abstractions}
\label{sub:spec_abstract}

A requirement or a specification is a statement which translates or expresses a need and its associated constraints and conditions where \cite{hall2020model}:

\begin{itemize}
\item \textbf{High-level requirements} regard the \textit{intent} of the system, rather than the goals it aims to achieve, independent of implementation details.
\item \textbf{Derived sub-requirements} or ``lower-level'' requirements result from design or implementation decisions necessary to satisfy a set of higher-level requirements. These sub-requirements can possess implementation detail, in addition to a more granular level of intent, which even further sub-requirements can be derived from.  
\end{itemize}

Higher-level requirements or specifications are often distinct from lower-level specifications through the allocation of further structure and behavior within a defined boundary to satisfy one or more higher-level requirements. That is, the lower-level the specification, the more well-defined the architectural and programming constructs become. Indeed, there would be more ambiguity and difficulty in defining higher-level specifications for code synthesis, as the algorithm would need to implicitly derive an internal set of ``lower-level'' specifications before synthesizing the corresponding code solution. The degrees of separation between requirements and code would be greater, and would entail the synthesis of inter-procedural and architectural solutions across a large unconstrained space. If a lower-level specification is provided with well-defined constraints, this not only restricts the possible solutions, but lowers the degrees of separation between the specification and the code required to be produced (e.g., one function). As previously noted, the current capabilities of synthesis methodologies are only able to tackle tightly specified, constrained problem instances or narrow tasks.

\subsection{Computational and State Reasoning}
\label{sub:comp_reasoning}
Beyond the specification abstraction level, certain tasks require more complex computational constructs and state reasoning. In this section, we outline a set of programming language-independent properties that would be practiced by developers at various degrees of expertise and thus would implicitly be expressed in natural language prompts and specifications. These include:

\begin{itemize}
\item \textbf{Variable Interdependencies:} understanding and tracking the state of more than one variable, their interdependencies and nesting, all possible permutations of the state, and the relationship between input and output parameters
\item \textbf{Temporal Reasoning \cite{abadiLamport80}:} as consideration of future and past program states including
\begin{itemize}
    \item Safety properties entailing that a defined “bad” state never occurs
    \item Liveness properties entailing progress towards a specific goal or state
\end{itemize} 
\item \textbf{Concurrency and Parallelism:} Correct and sound reasoning over computational interleavings (for various specification granularities). The code generation technique should be able to reason or synthesize solutions requiring the following properties:
\begin{itemize}
\item Absolute Fairness or impartiality: every process should be executed infinitely often \footnote{Note that the usage of ''fairness" in this section explicitly regards computational concurrency and parallelism, and not unjust treatment.}
\item Strong Fairness: every process that is infinitely often enabled should be executed infinitely often in a state where it is enabled
\item Weak Fairness: every process that is almost always enabled should be executed infinitely often
\item Mutual exclusion and atomicity when needed
\item Correct synchronization
\item Freedom from race conditions and data races
\end{itemize}
\item \textbf{Nondeterminism:} In computational theory, a nondeterministic algorithm can provide different outputs for the same input on different executions. Unlike a deterministic algorithm which produces only a single output for the same input even on different runs, a non-deterministic algorithm travels in various routes to arrive at the different outcomes. A very simple and common example of this is a random number generator.\footnote{A randomized algorithm is actually a probabilistic Turing Machine, but for practical intents and purpose it can be approximately considered non-deterministic given the determinism of real-world systems \cite{barrington2000lecture}.} A more advanced and extreme example is ML algorithms themselves.
\item \textbf{Hyperproperties \cite{clarkson2014temporal}:} Information-flow policies and cryptographic algorithms requiring observational determinism which requires programs to behave as (deterministic) functions from low-security inputs to low-security outputs, for example:
\begin{itemize}
\item Noninterference: when the outputs observed by low-security users are the same as they would be in the absence of inputs submitted by high-security users.
\item Declassification: programs that need to reveal secret information to fulfill functional requirements.
\item Information-flow: policies that permit leakage of information at restricted rates. This includes min-entropy, which quantifies the amount of information an attacker can gain given the answer to a single guess about the secret.
\end{itemize}
\end{itemize}

Additionally, we note to the reader that there are a number of specification-independent coding practices that must be exhibited to achieve the aforementioned computational and state reasoning attributes. Such coding practices have long been discussed by the genetic programming community \cite{koza1999genetic}, and we note the relevant properties to modern day synthesis techniques below:

\begin{itemize}
\item \textit{Code and parameterized reuse:} Model has the ability to automatically organize useful groups of steps so that they can be reused. This includes various kinds of modularity, complex data types and control structures, and the potential to generate or modify instances of modularity, data and control structures with different values.
\item \textit{Automatic determination of program architecture:} Model has the ability to automatically determine whether to synthesize subroutines, iterations, loops, recursion, and internal storage, and the number of arguments utilized by each subroutine, iteration, loop, and recursion.
\item \textit{Wide range of programming constructs:} Model has the ability to implement a diverse set of programming constructs that human developers find useful, including macros, libraries, typing, pointers, conditional operations, typed functions, etc.
\item \textit{Well-defined:} The ability to distinguish between what the user must provide and what the system delivers.
\item \textit{Wide applicability:} Model produces a satisfactory solution to a wide variety of problems from many different domains (e.g., embedded systems, web applications, console applications).
\end{itemize}

Indeed, such constructs are required by developers when solving for increasingly complex and higher-level specifications. Without them, it is unlikely that a code generation model can tackle increasingly complex specifications describing and requiring the computational and state reasoning attributes noted.

As previously noted, many of the attributes above regard implementation level design. Increasingly higher level specifications should not need to specify which programming constructs are required by implementation, and a code generation algorithm should be able to infer this instead. Indeed, familiarity with certain specifications or prompts can lead to very successful outputs, but Codex struggles to generalize under unique circumstances when given increasingly complex or higher-level specifications.

\subsubsection{Evaluation and Limitations}
A challenge for traditional code generation is that, in the absence of formal specifications, we rely on the assumption that user intent is captured sufficiently enough such that the accuracy and synthesis of a methodology are not compromised. This is difficult to assume for Codex given the unreliable (and uncategorized) nature of the training data. For example, one consequential word is often the difference between Codex producing correct or incorrect results. Other factors such as:
\begin{itemize}
\item the context of existing code by a user,
\item defined function and variable names,
\item existing comments and documentation by a user,
\item training data distribution, and
\item conciseness and length of prompt,
\end{itemize}
heavily affect Codex’s capabilities to synthesize optimal or correct solutions. It is thus difficult with absolute certainty to state if Codex is proficient in meeting the evaluation criteria outlined in Section~\ref{sub:spec_abstract} and Section~\ref{sub:comp_reasoning}. Finally, Codex has been primarily trained on Python, Javascript, Typescript, and Ruby codebases, languages that are associated with specific domains such as web, application, or ML development. Dynamically typed languages are not the typical choice for implementing systems requiring constructs such as concurrency or cryptography algorithms (as with C/C++). Codex may thus only be proficient at synthesizing domain solutions optimal for languages for which it has been trained on.

\begin{itemize}
\item \textit{Variable interdependencies}: Codex has demonstrated encouraging results when reasoning about two or three program variables or datastructures, including the relationship between input and output variables. However, when faced with inter-reasoning over four or more variable relationships, especially when given unique prompts, Codex struggles to deduce the relationship between the presented variables and the intended output of the function. This is despite the specifications provided being relatively short and not significantly high-level. We anticipate that unless the specification description appears fairly frequently within the training data, that Codex will continue to struggle with variable interdependencies beyond three or more variables.
\item \textit{Temporal reasoning}: For short and narrow specifications, Codex performs relatively well when prompted to enforce a safety property (e.g., no division by zero) or a liveness or termination condition (e.g., when to exit a program or loop). However, when prompted to synthesize more complex and unique specifications, Codex fails to produce any  or correct outputs. This was the case for specifications that were not particularly high-level, and included attempts to define design and programming constructs. If a prompt was a common exercise or problem, Codex was able to synthesize the intended results. 
\item \textit{Concurrency and parallelism}: Codex’s performance so far indicates poor output and large reasoning gaps when synthesizing code requiring use of concurrency at any level of specification abstraction. All results thus far did not correctly synthesize solutions requiring fairness, atomicity, and/or synchronization.
\item \textit{Nondeterminism}: Codex performed well for small constrained tasks such as random number generation. For more complex tasks such as building ML models, Codex demonstrated productive results as it was able to effectively generate boilerplate ML code, especially for common portions of well used codebases (e.g., MNIST loading code). Although Codex did not always generate the correct outputs for nuanced or uncommon prompts, it synthesised enough boilerplate code that could be easily tweaked by a user to correct for any inaccuracies. This has the potential to accelerate ML model building.
\item \textit{High-level specification and automatic determination of architecture}: Codex use and output is most optimal when specifying problems that can be constrained to one function or module-level implementation. For a module, the capacity for Codex to synthesize correct code and programming constructs is largely correlated with the data available, rather than the level of abstraction or conciseness a specification may be written at. However, if one were to define specifications that must be solved across multiple modules with automatic determination of program architecture, Codex would struggle to synthesize such requests. This entails that high-level systems specifications (e.g. requirements for an aircraft) are currently beyond the scope of Codex’s capabilities. However, we have observed in some instances that Codex synthesizes “getter” helper functions. Although simple, this may be an indication to potential interprocedural synthesis that would tackle system-level specifications.
\item \textit{Hyperproperties}: Given the limitations and shortcomings of Codex noted above, it's challenging to devise a synthesis prompt that would satisfy non-interference or information-flow policies. That is, Codex does not possess the capabilities to synthesize building blocks that could allow for synthesis of cryptography algorithms with complex hyperproperties.
\end{itemize}
%\vspace*{-2.5mm}
We note that Codex does not guarantee correctness or soundness of any solutions produced. Indeed, we have observed that Codex can often recommend syntactically incorrect code and functions, variables, and attributes that are undefined and not within the scope of the codebase or libraries used. It is also not uncommon that Codex recommends modules from libraries which have not been declared or imported (by the user or Codex itself). Finally, despite no implementation limit to the length of the prompt, Codex struggles to parse through increasingly long specifications, likely reflective of comments structures within the training data.

The evaluation of the above metrics provided a generative capabilities baseline used to inform both the Codex evaluation in~\cite{chen2021evaluating}, and our risk assessment below. A capabilities evaluation being carried out {\it prior} to a risk assessment may seem counter-intuitive. However, traditional risk assessments require implicit assumptions and knowledge regarding a prospective system's capacities, limitations, and failure modes (which in turn inform possible harms a system may pose). In the case of code synthesis LLMs, and more generally LLMs, these capabilities and failure modes are not yet fully understood. The evaluation of the above metrics provides said generative capabilities baseline needed for code synthesis models exceeding previous state of the art.

\section{Hazard Analysis and Risk Assessment}\label{HazardAnalysis}

In this section, we describe the hazard analysis and risk assessment approach taken at OpenAI for systems involving Codex-like models as components. Our reference point for consideration is the Codex API that serves outputs to users, though our analysis approach is relevant to many other kinds of systems as well. Our risk assessment considers the risks attached to generative uses of these models in consideration against their generative capabilities as evaluated in Section \ref{Evaluation}.

There are numerous approaches, techniques, and levels of rigor in carrying out a hazard analysis and a risk assessment, and we refer to existing literature for further detail \cite{leveson2016engineering}. Our approach is reminiscent of a preliminary System Hazard Analysis (SHA) that subsumes a further categorization and prioritization of the hazards across each “subsystem”\footnote{Work in~\cite{weidinger2021ethical} proposes a taxonomy of six risk categories, with the code synthesis LLM risks being derived from our work initially noted in~\cite{chen2021evaluating}. A more exhaustive list is provided further below. Our risk assessment additionally differs in that we distinguish between hazard sources and risk categories corresponding to each source. This leads to our risk categories to be partitioned between those regarding the construction of the model, versus the application of the model itself.} (i.e., Subsystem Hazard Analysis). The SHA-like approach ensures coverage of a wide scope of hazard sources including:
\begin{itemize}
\item Applications (E.g., Human health, Opportunity and Livelihood, Social and Political Cues, Microtargeting, Integrations to Safety-Critical Systems, Government \& Civics)
\item Alignment (which, here, we interpret as the degree to which the behavior of the AI does or does not accord with user intentions; misaligned AI may produce unsafe behavior) \cite{chen2021evaluating, askell2021general, leike2018scalable}
\item System Design and Implementation (e.g., UX/UI, Documentation, Requirements, Data Provenance, Validation)
\item Regulatory and Legal Oversight (e.g., Intellectual Property, Export Control, Data Privacy \& Rights)
\item Defense and Security
\item Economic and Environmental Impacts
\end{itemize}
%
%\footnote{Note that we distinguish between hazards that may arise from construction of the model, versus the application and output of the model itself under (e.g., “Application” and “Alignment”).}
Risk assessments frameworks require a defined set of Hazard Severity Categories (HSC). However, the standard definitions utilized (e.g, \cite{dod2012mil}) across all industries are not sufficient to accommodate for novel safety issues that LLMs and their applications pose. In Table ~\ref{tab:hazard_cat}, we thus propose a novel set of HSC associated with the use of language model APIs, supported by a set of defined harms and losses (see Table ~\ref{tab:losses}) that may be used as foundations for safety efforts for all language models. We believe this expansion of the standardized definitions of HSC will not only bolster the use of traditional hazard analysis practices within the ML community, but will allow those industries that utilize hazard analysis to appropriately consider novel harms posed by all uses of LLMs (e.g., GPT-3).

As in any traditional risk assessment, hazards are then prioritized to recognize which hazards are of greatest concern through a defined risk model. We use the standard Hazard Risk Index (HRI) as a metric to note the initial risk perceived for each hazard. Typically an HRI is based on the product of the probability of events against their severity, but given the novelty of Codex-like models and systems built around them, quantitative data and analysis is not currently always possible to achieve. A quantitative probability guide with corresponding qualitative metrics was used based on \cite{dod2012mil} for hazard probabilities (i.e., Frequent (A), Probable (B), Occasional (C), Remote (D), Improbable (E)). When we performed our hazard analysis for the Codex API, we used the results from evaluations in Section \ref{Evaluation} to inform our estimates of hazard probabilities. The cross product of the above HSC and qualitative hazard probability levels are then used to form the HRI in Table ~\ref{tab:hri}.

{\centering
\begin{table}[t]
% \small
\begin{tabular}{|l|r|>{\raggedright}p{0.7\textwidth}|}
\hline
\textbf{Description} &
  \textbf{Category} &
  \textbf{Definition (Mapped to Table 2)} \tabularnewline \hline
Catastrophic &
  \cellcolor[HTML]{CC0000}1 &
  Death, permanent total disability, direct harm, system loss, or irreversible significant environmental impact. \tabularnewline \hline
Critical &
  \cellcolor[HTML]{E69138}2 &
  Permanent partial disability, injuries, incitement, manipulation, radicalization, or discriminatory harm that may result in hospitalization of multiple people. Cause of consequential error to many individuals or reversible significant environmental impact. \tabularnewline \hline
Major &
  \cellcolor[HTML]{F1C232}3 &
  Injury or cause of consequential error to a few individuals, or reversible moderate environmental impact. \tabularnewline \hline
Minor &
  \cellcolor[HTML]{6AA84F}4 &
  Injury or cause of consequential error not resulting in any long term harm, or minimal environmental impact. \tabularnewline \hline
\end{tabular}
\caption{\label{tab:hazard_cat}Hazard Severity Categories associated with the use of language model APIs.}
\vspace{-8mm}
\end{table}
}
{\centering
\begin{table}[!htbp]
\resizebox{\textwidth}{!}{%
\begin{tabular}{|r|>{\raggedright}p{0.45\textwidth}|>{\raggedright}p{0.65\textwidth}|}

\hline
\textbf{ID} & \textbf{Loss} & \textbf{Example/Rationale} \tabularnewline 
\hline
\textit{L1} & \textbf{Direct harm.} Information created by or provided via the API causes or contributes to risk of physical, emotional, psychological injury, damage to property, or damage to the environment; denial of consequential services; infringement on human rights; or the erosion of social \& democratic structures. & For example, includes situations where: \begin{itemize}[leftmargin=0.5cm]
    \item the API generates abusive content that causes someone to suffer,
    \item or exacerbates psychological harm experienced by end users (eg encouraging suicide in a therapy setting, or feeding addictive behaviors),
    \item or where the API is involved in controlling a physical system to damage itself or the world around it.\end{itemize} \tabularnewline
\hline
\textit{L2} & \textbf{Incitement, manipulation, or radicalization.} Information created by or provided via the API is a significant cause for someone to commit harm against others or themselves, property, or the environment, or otherwise drives people to participate in extremist acts or groups.                             & For example, includes situations where:\begin{itemize}[leftmargin=0.5cm]
\item the API persuades an end user to commit a direct harm, eg by affirming or suggesting violent pathological behavior in a therapy setting,
\item or where the API is involved in recommendation engines that facilitate radicalization by directing users to hateful content.\end{itemize}  \tabularnewline 
\hline
\textit{L3} & \textbf{Discriminatory harm.} Information created by or provided via the API is a contributing factor in the perpetuation of systemic harms against any group, including an oppressed, marginalized, or underrepresented group of people.                                                                             & For example, includes situations where:\begin{itemize}[leftmargin=0.5cm] \item the API invisibly discriminates in evaluating loan or job applicants, eg by implementing redlining-like policies in ways not understood by human application managers,
\item or where the API generates racist, sexist, or otherwise discriminatory content,
\item or where the API underrepresents some groups or people in a harmful way
\end{itemize} \tabularnewline \hline
\textit{L4} & \textbf{Causing consequential error.} Information created by or provided via the API causes people or institutions to make errors in judgment, for example through false beliefs or faulty premises, that directly or indirectly have adverse impacts on quality of life.                                             & For example, includes:\begin{itemize}[leftmargin=0.5cm]
\item damage to the information environment that people rely on for personal, political, technical, or medical decisions,
\item or uses of the API that result in people experiencing loss of opportunity or being denied just access to an important resource or service,
\item or uses of the API in high stakes decision-making tools that are founded on unscientific premises, eg a tool that purports to detect criminality based on a person's appearance or writing style.
\end{itemize}

In a broad way, covers misinformation, but only misinformation that leads to harm (e.g. giving the wrong answer when asked who a celebrity is currently dating is not a concern).  \tabularnewline
\hline
\end{tabular}}
\caption{\label{tab:losses}Losses Definitions}
\vspace{-8mm}
\end{table}}

Finally, in Table ~\ref{tab:risk_examples} we provide an illustrative view of our final risk assessment approach with a sample simplified list of hazard sources, descriptions, and controls identified for the Codex API. The preliminary HRI for each hazard help us understand how risks compare to each other, and whether a given hazard is worth controlling. We note that risk assessments should be carried out by a multidisciplinary team with backgrounds in safety, policy, security, engineering, and law to ensure comprehensive coverage of possible hazards and risks.

In the next sections, we outline some of the most notable and urgent risks identified in carrying out this risk assessment against the aforementioned Codex performance baseline, followed by a set of mitigations that are applicable to all large language code synthesis models.

{\centering
    \begin{table}[]
% \small
\begin{tabular}{|l|l|}
\hline
\rowcolor[HTML]{F7CB4D} 
\textbf{HRI}              & \textbf{Risk Decision Criteria} \tabularnewline \hline
\cellcolor[HTML]{CC0000}1A, 1B, 1C, 2A, 2B, 3A & Unacceptable; stop operations and rectify immediately.           \tabularnewline \hline
\cellcolor[HTML]{DE7E2B}1D, 2C, 2D, 3B, 3C & Undesirable; upper-management decision to accept or reject risk. \tabularnewline \hline
\cellcolor[HTML]{ECB727}1E, 2E, 3D, 3E, 4A, 4B & 
 Acceptable with management review.                               \tabularnewline \hline
\cellcolor[HTML]{599B3E}4C, 4D, 4E & Acceptable without review.               \tabularnewline \hline
\end{tabular}
\caption{\label{tab:hri}Hazard Risk Index}
\vspace{-8mm}
\end{table}}
{\centering

\begin{table}[]
\renewcommand\arraystretch{2} % increase spacing between rows
\resizebox{\textwidth}{!}{%
\begin{tabular}{|>{\raggedright}p{0.12\textwidth}|
>{\raggedright}p{0.2\textwidth}|
>{\raggedright}p{0.2\textwidth}|
>{\raggedright}p{0.2\textwidth}|
>{\raggedright}p{0.2\textwidth}|
>{\raggedright}p{0.2\textwidth}|
>{\raggedright}p{0.1\textwidth}|
>{\raggedright}p{0.15\textwidth}|}% Please add the following required packages to your document preamble:
% \usepackage[normalem]{ulem}
\hline
\rowcolor[HTML]{F7CB4D} 
\textbf{Hazard Source} &
  \textbf{Hazard Description} &
  \textbf{Trigger Events} &
  \textbf{Effects} &
  \textbf{Hazard Risk Index (HRI)} &
  \textbf{Hazard Control(s)} &
  \textbf{Effect of Control on HRI} &
  \textbf{Verification of Control} \tabularnewline \hline
\rowcolor[HTML]{FFFFFF} 
\textbf{Application - Integrations to Safety-Critical Systems} &
  Unfettered capabilities of state actors or others to build safety-critical systems &
  Providing a high-level specification that defines the intent or bounds of an aerospace or weapons system, for which Codex successfully synthesizes some aspects of functionality. &
  Malicious state actors or political groups and entities building systems with more ease that lead to death or harm of both civilians and military personnel. &
  1E - Codex is currently not capable of synthesizing code beyond tightly specified, constrained problem instances or narrow tasks. & 
  \begin{itemize}[wide=0pt, leftmargin=*, nosep, before =\vspace*{-0.5\baselineskip}]
      \item Rate limiting 
      \item Limit generation of nested/helper functions
      \end{itemize}
% \begin{itemize}[leftmargin=*]
%       \item Rate limiting 
%       \item Limit generation of nested/helper functions
%       \end{itemize}
      &
  3E &
  \begin{itemize}[wide=0pt, leftmargin=*, nosep, before =\vspace*{-0.5\baselineskip}]
      \item Continuous evaluation of Codex’s capabilities as part of product life cycle 
  \end{itemize} \tabularnewline \hline
\rowcolor[HTML]{FEF8E3} 
\textbf{Application - All usages} &
  Codex generates completions that encode bias in ways that disproportionately harm or benefit different groups. (This could be exacerbated if completions are seen to be "standard" or correct approaches.) &
  Codex used to  generate code to perform classification along the lines of gender or other sensitive characteristics such physical or mental attributes, race, nationality, socio-economic status, etc. &
  Codex suggests code that assumes binary gender, resulting in an application that misgenders people and reinforces assumptions around binary gender.\newline Codex exacerbating false and harmful stereotypes against marginalized groups. &
  2B - This is an instance in which distribution of harm is a critical consideration, in addition to severity, probability, and frequency. &
  
    \begin{itemize}[wide=0pt, leftmargin=*, nosep, before =\vspace*{-0.5\baselineskip}]
  \item Usage and access policies 
  \item Blocking completions 
  \item Documentation of model characteristics and limitations 
  \item Data provenance and curation to mitigate against such harms. 
  \end{itemize}
  &
  2C &
    \begin{itemize}[wide=0pt, leftmargin=*, nosep, before =\vspace*{-0.5\baselineskip}]
  %\noindent{\begin{itemize}[leftmargin=.3cm]
  \item Red teaming exercises
  \item Continuous evaluation of CodeGen's capabilities as part of product life cycle
  \end{itemize}
  \tabularnewline  \hline
\rowcolor[HTML]{FFFFFF} 
\textbf{Alignment} &
  Codex produces code with bugs when prompted with code that includes bugs (even those that may be subtle/accidental on the part of the coder) &
  A coder is using Copilot to make some improvements to a codebase and enters a prompt with a bug, which Copilot completes with further defects &
  Copilot suggests vulnerable code, resulting in an unsafe codebase that compromises the security and privacy of downstream users. &
  2B &
    \begin{itemize}[wide=0pt, leftmargin=*, nosep, before =\vspace*{-0.5\baselineskip}]
  \item Application of Static Analysis and Security tools 
  \item Targeted training to block malicious suggestions
  \item Documentation of model characteristics and limitations 
  \end{itemize}
  &
  2C &
    \begin{itemize}[wide=0pt, leftmargin=*, nosep, before =\vspace*{-0.5\baselineskip}]
  \item Red teaming 
  \item Human Evaluation to gauge alignment outcomes 
  \end{itemize}
  \tabularnewline  \hline
\end{tabular}}
\caption{\label{tab:risk_examples}Risk Assessment Framework}
\vspace{-8mm}
\end{table}}

\section{Risk Assessment Outcome}
\label{Risk}
In this section we provide a summary of the more pressing hazards based on the potential hazard sources noted in Section 3.\footnote{Due to space constraints, we are not able to provide an exhaustive list of all hazards identified and their corresponding analyses. However, we hope that our example framework and most pressing risks identified allow those constructing code synthesis LLMs to appropriately assess hazards and risks specific to their models.} We emphasize application hazards given the uncertainty of what deploying an application utilizing Codex-like models would entail societally, economically, and politically. 

\subsection{Application}
\textit{Discrimination, Fairness, and Bias}
\begin{itemize}
\item Potential inline generation features being utilized in an open-ended manner permitting general, non-code usage of language model capabilities, mirroring what has been found in the case of other language
models trained on Internet data \cite{bender2021dangers, blodgett2020language, abid2021persistent, brown2020language}.
\item Generation of completions that encode bias in ways that disproportionately harm or benefit different groups (this could be exacerbated if completions are seen to be "standard" or status-quo approaches).
\item Inadvertently producing biased and discriminatory code if prompted to by comments or auto completion.
\end{itemize}
\textit{Security}: Inadvertently suggesting malicious or vulnerable code (including library use) that compromise the safety or security of the application being developed, or the system which it operates on, including safety-critical systems. A concurrent study has demonstrated these results further using Github's CodeQL~\cite{githubQL}.\\
\textit{Safety-Critical}
\begin{itemize}
\item Use of synthesis to build or infer information pertaining to safety-critical systems. This may provide malicious laymen the capabilities to construct (through inference or direct code generation) complex aerospace, nuclear, or defense technologies that give them unfettered capabilities of state actors that pose threats to civilians. For the current implementation of Codex, this risk is not high given its noted limitations, but may increase as the model advances.
\item Accelerating use of neural network model development, including reducing the cost of disinformation operations, deep fakes, surveillance, facial recognition, etc.
\end{itemize}
\vspace*{-5mm}
\subsection{Alignment}
%\textit{Alignment}
\begin{itemize}
\item Producing code or comments with mistakes, when prompted with code or comments that include mistakes or bugs (even those that may be subtle or accidental on the part of the coder).
\item Suggesting solutions that superficially appear correct but do not actually perform the task the user intended, negatively affecting productivity and learning of novice programmers.
\end{itemize}
\vspace*{-5mm}
\subsection{System Design and Implementation}
\begin{itemize}
\item \textit{Requirements and Documentation}: Lack of requirements or understanding of the model’s or API’s features and limitations, including UI misleading users to have overconfidence in the AI's ability.
\item \textit{UI/UX}: UI is inaccessible to marginalized communities.
\item \textit{Accuracy and Performance}: Overreliance and over-trust on the model to generate mission-critical output (e.g., documentation or comments), leading developers to miss implementation and safety relevant details that would otherwise be observed by manual processes. (Casually, we refer to this as ``falling asleep at the wheel.'')
\end{itemize}
\vspace*{-5mm}
\subsection{Regulatory and Legal Oversight}
\begin{itemize}
\item Ambiguous legal liability for model creators, customers, and end-user of inadvertent use of Intellectual Property or incorrect use of licensed code (e.g., General Public License)~\cite{chen2021evaluating}.
\item Foreign made items incorporating 25\% or more of controlled U.S.-origin content are potentially subject to the Export Administration Regulations (EAR) for purposes of export or reexport \cite{reexportguidance}. Given that this is applicable to software systems, model usage may fall under export control.
\end{itemize}
\vspace*{-5mm}
\subsection{Economic and Environmental Impacts}
\begin{itemize}
\item Synthesis produces code and comments, which are key components of some software development jobs, and thereby increases potential of displacement of certain jobs.
\item Access to synthesis tools and the associated productivity gains serve to concentrate power and exacerbates inequality, constricting economic growth.
\item Using synthesis tools requires certain amount of technological literacy, hardware, and an internet connection, implicitly excluding the most economically vulnerable from direct economic benefits and widening existing economic opportunity gaps.
\item Excluding individuals and businesses from access to synthesis tools based on the country they live in risks drives economic inequities across countries, and granting it in a way that is not inclusive within a certain country exacerbates inequality.
\item Synthesis features are used to generate code for application with environmental impacts, exacerbating environmental hazards. Synthesis tools themselves are energy-intensive due to compute requirements, potentially causing environmental harm via compute supply chains and non-renewable energy consumption.
\item Synthesis disproportionately benefits or harms a certain subset of software developers, in a way that exacerbates demographic disparities within the field (e.g. disproportionately affecting front end software development, which tends to be more demographically diverse than other subsets of the field).
\end{itemize}

\section{Hazard Controls and Mitigations}
\label{HazardControls}
In this section, we describe a wide range of hazard controls and mitigations that can be implemented to eliminate or reduce the risks identified in the hazard analysis. As in the hazard analysis, some of these mitigations are intended for API systems that enable users to query Codex-like models with arbitrary prompts, and different mitigations may be appropriate for other kinds of systems. Given that prioritization for which mitigations to implement should depend on system-specific factors, including local costs and trade-offs and the level of capabilities for the specific models involved, we do not recommend a specific prioritization among this space of potential mitigations. However, when system designers choose mitigations to implement, an appropriate basis for selection is the ALARP principle, a known approach that states that the residual risk of a system shall be as low as reasonably practicable \cite{britain1974health}. The key factor to ALARP is the emphasis to balance the realized safety benefits to the actual costs to implement.
% As with all risk assessments, once a prioritized set of hazards is determined, we implement hazard controls that can either eliminate or reduce the risk of the hazard from occurring. In defining our mitigations, we adopt the ALARP principle, a known approach that states that the residual risk of a system shall be as low as reasonably practicable \cite{britain1974health}. The key factor to ALARP is the emphasis to balance the realized safety benefits to the actual costs to implement.

We partition our mitigations into two categories:
\begin{itemize}
\item Plausible and Immediate: These are technologically feasible solutions that those constructing code synthesis LLMs may have the capability to implement directly today.
\item Long Term: These are solutions that would contribute to ensuring the safety of code synthesis LLMs, but which may be open research problems, or require significant resources invested over the entire life cycle of the system.
\end{itemize}

We note that although “plausible” solutions may have the most immediate impact given their accessibility, they do not reflect the severity of the risks to which they are applied to. That is, these are mitigations which can be tackled immediately and more often than not, may still lessen the hazard of even the highest risks, even those that require longer-term solutions.

\subsection{Plausible and Immediate Mitigations}
\subsubsection{Documentation and Communication Channels}
Documentation can help mitigate potential harms posed by the use of code generation systems by communicating acceptable uses of the technology and potential safety risks associated with particular uses. %By defining best (and legal) practice, we encourage developers to exercise good judgment. 
%
% Here we distinguish between “end-users” who are end-users of applications and technologies built with Codex and “workers” that either use the model in their own coding or work for firms that use the model. Furthermore, documentation and disclaimers must be provided that are agreed to prior to use by both end-users, and workers that clearly communicate the following:
It would be helpful to provide documentation to direct users of the code generation system as well as downstream end users of the applications and technologies built with it. Documentation and disclaimers might include:
\begin{enumerate}
\item the characteristics, limitations and potential shortcomings of the code generation models, possibly in the format of a model card \cite{mitchell2019model},
\item that a decision, content, advice or outcome is the result of an algorithmic decision,
\item the amount of data that is logged and collected by the code generation system (i.e. to train future models, study worker productivity, etc.),
\item the level of specialized knowledge (i.e., expertise in software development) required to operate the code generation system to be able to distinguish between correct or incorrect solutions,
\item that the model does not guarantee any sound or complete results regarding synthesis, generation, summarisation, or other uses,
\item whether the code generation system has been certified for use in generating safety-critical solutions, and if so, in what domains, on the basis of what evidence, and the required level of qualifications for users,% n the case that safety or mission critical domain is approved for use, it should also be noted that only expert users (e.g., in safety, software, and verification) should have access where Human-In-Loop verification and validation must be carried out,
\item information about applicable laws and regulations that may apply to software engineering products created with the assistance of the code generation system (e.g., foreign-made items incorporating 25\% or more of controlled U.S.-origin content are subject to EAR \cite{reexportguidance}).
\end{enumerate}

Because models with Codex-like capabilities are comparatively new and we cannot yet predict the full range of their capabilities and impacts, we also recommend the creation of channels for users and impacted stakeholders to engage directly with model creators to raise concerns or report acceptable use violations.

\subsubsection{Product and API}
When the system encapsulating a code generation model is an API or similar, many options are available for the system designers to implement mitigations at the API level by restricting the space of possible user actions. We expect the following mitigations to have broad relevance:

%A model being encapsulated within a product or API has significant impacts on the ability to mitigate for hazards across the board, by actively restricting the feature space to what can be accomplished. No model should be released as a product or API without some exploration of the following:
\begin{itemize}
\item \textbf{Rate-limits and operational constraints} (e.g., restricting the number of allowed API requests per unit time, and restricting the number of model outputs per API request)
\begin{itemize}
\item Before a high degree of confidence is established that all potential hazards have been mitigated against, rate-limits may be seen as a primary tool to derisk applications and users, even if such conservatism inhibits (non-malicious) application development.
\item Collect statistics on normal usage volume to use as a base metric, and identify usage volume levels that would be considered anomalous and might indicate malicious use. %Utilizing monitoring to quantitatively understand non-anomalous use of the model as a base metric to define the operational bounds of the product or API released. This will require requesting usage information from users.
\end{itemize}
\item \textbf{Filtering, flagging, and monitoring}
\begin{itemize}
\item \textbf{Blocking a subset of outputs:} Code generation systems should not suggest language that is inherently harmful or toxic. In between when model outputs are generated and when the API serves them to a user, they can be checked against word filter lists or other classifiers to determine if they are acceptable to serve; unacceptable outputs can be blocked. %Adopting from existing LLM monitoring and content filtering would accelerate progress on this.
\item \textbf{Gate completions of particular tokens by user input:} Ensure some high threshold of confidence that the user’s intention is fully captured before offering a suggestion, especially where it may encode social or cultural values. For example, it would not always be contextually appropriate to autocomplete ``f'' to “female”, but it would be more reasonable if the user typed “fem” first.
\item \textbf{Organization or user level filter:} Enable organizations or users using an API at the enterprise level to define a list of filtered tokens. Filtering at the org level would potentially require a more robust permissioning systems.
\item \textbf{Inform ongoing efforts with know your customer (KYC):} Construct monitoring probes to build an understanding of how end-users are using API or product in order to tailor classifier filters for identified hazards in the future.
\end{itemize}
\item \textbf{Eliminate completions in contexts where behavior and/or performance are knowingly uncertain, unsound, insecure, and/or unsafe} 
\begin{itemize} 
\item Leverage existing code analysis capabilities (e.g., Github’s Semmle) to analyze completions in context with programming language utilized
\begin{itemize}
\item \textbf{Syntactic analysis:} Code compliance checkers (e.g., PyPI black) can build confidence in the quality of the code through identifying poorly constructed code and syntactic non-conformance. This entails re-consideration of token completion to align with an Abstract Syntax Tree (AST) structure of the programming language at hand to optimize use with code compliance checkers.
\item \textbf{Semantic analysis:} Using formal verification techniques to understand behavior of proposed synthesized completions. Usage of such tools would be most effective on synthesis completion at a module-level. This would help eliminate unsafe language completions (e.g., buffer overflow) that may lead to safety and security vulnerabilities. Note this would not be viable to dynamically typed languages, which are not amenable to verification.
\end{itemize}
\item Detect whether users are attempting to subvert the intended usage of code generation models with adversarial prompts that may unlock open-ended generative language behavior (e.g. trigger the model to enter a conversational dialogue mode). This detection may be implemented by checking for the absence of code in model outputs, either via classifiers, code analysis tools, or naive regex and pattern matching. % In the absence of classifiers and code analysis tools, use naive regex and pattern matching to determine when users are attempting to circumvent bounds of Codex or other models that build on LLMs (e.g., GPT-3) in an open-ended manner permitting general, non-code usage of language model capabilities
\end{itemize}
\item \textbf{Build UX with an eye toward safety concerns.} Design elements within the user interface may prevent a user “falling asleep at the wheel” and inadvertently accepting bad code suggestions. These may include:
\begin{itemize}
\item Marking or highlighting generated code (or, with a robust classifier, potential mistakes)
\item Adding delays between calls to the model to encourage users to review generated code and discourage anomalous behavior
\end{itemize}
\end{itemize}

\subsubsection{Data Provenance}
Construction of infrastructure for the data pipeline for data-readiness and understanding the assumptions and limitations of data utilized. This includes evaluating the real-world data for quality, validity, and availability and its effects on model performance and outputs. This includes:
\begin{itemize}
\item Oversight mechanisms for data collection, storage, processing
\item Understanding minimal use or implications of potentially sensitive or personal data and compliance with existing laws (e.g., California Privacy Protection)
\item That the persons are qualified and required to access the data, with oversight mechanisms to log when, where, how, by whom and for what purpose data was accessed.
\end{itemize}

In terms of regulatory or privacy violations, following industry best practices (equivalent to ISO 8000 Part 140 and ISO 25012) to ensure that data is accurate, complete, credible, consistent, confidential, timely, and traceable. This includes addressing:
\begin{itemize}
\item Assessment of whether there is a need for additional data, for example to improve accuracy or to eliminate bias
\item Ensuring use of diverse datasets and consideration of representation to ensure alternative perspectives or viewpoints are included, or if harmful and toxic subsets require removal
\item Identifying training or test data for categories of interest be identified, when necessary, for auditability
\end{itemize}

\subsection{Long-Term Mitigations}
\subsubsection{ML Architecture and Implementation}
\begin{itemize}
    \item Adapt the Codex model architecture and API implementation to output tokens that align with programming language syntax or ASTs in order to allow model outputs to be more amenable for use for security and safety techniques. Deriving from programming language synthesis techniques may lead the way on how this can be carried out ~\cite{shinSynthesis,drewsSynthesis,Brockschmidt2019GenerativeCM, PGL-010,PGL-049}. Consideration of programming languages' syntax and semantics within an ML model would further allow us to distinguish between programming language and natural language completions, allowing us to constrain outputs or prompts that are attempting to tap into open-ended language model capabilities.
\end{itemize}

% \begin{itemize}
% \item Adapting the Codex model architecture and API implementation to output tokens that align with programming language syntax or ASTs in order to allow API outputs to be more amenable for use for security and safety techniques. Deriving from programming language synthesis techniques may lead the way on how this can be carried out. Such an architecture would further allow us to distinguish between programming language and natural language completions, allowing us to constrain outputs or prompts that are attempting to tap into GPT-3 like capabilities.
% \end{itemize}
\subsubsection{Fine-tuning, model re-training, and classifiers}
\begin{itemize}
\item Fine-tuning on a small but curated datasets can help improve language model behavior against discriminatory and biased outputs and have a larger impact as model size increases~\cite{PALMS}.
\item Blacklist and remove libraries from training data that allow for the acceleration of hazards including:
\begin{itemize}
\item cost reduction of disinformation operations, deep fakes, surveillance, facial recognition
\item security and zero-day exploits that may enforce application and system harm
\end{itemize}
\item Train classifiers to detect use of circumventing blacklisted code bases (e.g., through aliasing) and discriminatory or biased behaviour
\item Train models to help users create better specifications, e.g., by asking about areas of the specification that are unclear. We do not believe that enforcing any type of formal specifications would be beneficial to resolve ambiguities in prompts, but the models themselves may be able to assist users in determining their intentions.
\item Build classifiers for potential malicious use cases (e.g., when a malicious user might ask the model to write a SQL injection attack) and don’t serve completions on malicious suggestions.
\end{itemize}

\subsubsection{User and worker rights}
\begin{itemize}
\item A responsible vulnerability disclosure program as well as a bias bounty program could bring users and workers into future hazard and safety discussions.
\item Research compensation schemes for user data later used for training.
\end{itemize}

\subsubsection{Economic impacts}
\begin{itemize}
    \item Conduct research to understand the economic impacts of code synthesis LLMs and to develop tools to forecast impacts of future models. This includes collaboration with external researchers in developing partnerships that would allow us to understand the labor market impacts of code generation models and how these impacts may translate into safety risks and hazards.
\end{itemize}

The scale at which each hazard control or mitigation reduces the HRI for risks identified is still an open question, as these methods must be qualitatively or quantitatively evaluated overtime to determine a new HRI for each hazard (especially considering long-term mitigations). Despite the novelty of code synthesis LLMs, deployment of these mitigations will still lessen the hazard of even the highest risks. Verification and monitoring capabilities of these mitigations must thus be in place to ensure that hazards have been sufficiently controlled.

\section{Conclusive Remarks and Looking Forward}
\label{Conclusion}
With the advent of Codex and a high likelihood of more powerful models in the future, it’s necessary to evaluate code generation beyond toy function synthesis examples and assess the capabilities against human ability. Additionally, it has not been clear how to measure or gauge increasing levels of capabilities between model sizes and architectures. In this paper, we propose a novel evaluation framework of code synthesis LLMs, that aids in determining the capacity of advanced code generation techniques against the complexity and expressivity of specification prompts, and the models' capability to understand and execute them relative to human ability. This analysis underpins an outlined a hazard analysis framework constructed to uncover hazards or safety risks Codex may impose technically, socially, politically, and economically.

Through our evaluation and hazard analysis, we outline the pressing hazards identified applicable to all code synthesis LLMs, followed by a set of hazard controls and mitigations model creators should always consider when building novel code synthesis platforms. While we focus here on model capabilities, we emphasize that model evaluation should be conducted on an ongoing basis as part of safe development and deployment, including evaluation of performance in specific contexts of use and in the real world. 

\bibliographystyle{ACM-Reference-Format}

\bibliography{citations}

\end{document}